\documentclass[aps,prd,paper,superscriptaddress]{revtex4}
\usepackage{amsmath,amssymb,epsfig,natbib,bm,hyperref}
\usepackage{boxedminipage}
\def\be{\begin{equation}}
\def\ee{\end{equation}}
\def\bea{\begin{eqnarray}}
\def\eea{\end{eqnarray}}

\begin{document}

\title{Efficient Cosmological Parameter Estimation with Hamiltonian Monte Carlo}
\author{Amir Hajian}
\email[]{ahajian@princeton.edu} \affiliation{ Department of Physics,
Jadwin Hall, Princeton University,   PO Box 708, Princeton, NJ
08542. }\affiliation{ Department of Astrophysical Sciences, Peyton
Hall, Princeton University, Princeton, NJ 08544. }

\date{August 31, 2006}
\begin{abstract}
Traditional Markov Chain Monte Carlo methods suffer from low
acceptance rate, slow mixing and low efficiency in high dimensions.
Hamiltonian Monte Carlo resolves this issue  by avoiding the random
walk.  Hamiltonian Monte Carlo (HMC) is a Markov chain Monte Carlo
(MCMC) technique built upon the basic principle of Hamiltonian
mechanics. Hamiltonian dynamics allows the chain to move along
trajectories of constant energy, taking large jumps in the parameter
space with relatively inexpensive computations. This new technique
improves the acceptance rate by a factor of $4$ while reducing the 
correlations and boosts up the
efficiency by at least a factor of $D$ in a $D$-dimensional
parameter space. Therefor shorter chains will be needed for a
reliable parameter estimation comparing to a traditional MCMC chain
yielding the same performance. Besides that, the HMC is well suited
for sampling from non-Gaussian and curved distributions which are
very hard to sample from using the traditional MCMC methods. The
method is very simple to code and can be easily plugged into
standard parameter estimation codes such as \texttt{CosmoMC}. 
In this paper we demonstrate how the HMC can be efficiently 
used in cosmological parameter estimation.

\end{abstract}

\maketitle

\section{Introduction}

The wealth of information in the high quality data of the recent
Wilkinson Microwave Anisotropy Probe (WMAP) satellite experiment
combined with other recent advances in observational cosmology has
led the field of cosmology into the new era of precision cosmology
\cite{wmap3yr}, \cite{Tegmark:2003ud}. These information are used to
constrain the cosmological models, by putting estimates and error
bars on various quantitative parameters of the cosmological models.
Markov Chain Monte Carlo methods are the most popular techniques of
parameter estimation in cosmology. These methods first introduced in
the astrophysical context by \cite{Christensen:2001gj}. Since then
MCMC have been used in several problems in cosmology. Standard
packages were made publicly available for cosmological parameter
estimation including \texttt{CosmoMC} \cite{Lewis:2002ah},
\texttt{Analyze This!} \cite{Doran:2003ua}.  MCMC techniques have also been applied to the gravity wave
data analysis \cite{Cornish:2005qw}, \cite{Cornish:2006ry} and to modelling extrasolar planet data.

The MCMC methods are much faster than the grid-based ones but even
with the currently available data a naive MCMC algorithm takes a
long time to converge and estimating cosmological parameters
specially those outside the LCDM is a challenging task. This problem
is much more important for the future data of the upcoming
experiments like ACT \cite{ACT} and Planck \cite{Plank} that will
map the CMB sky on small scales (large $l$). It seems necessary to
design new methods and techniques to speed up the parameter estimation
in a more efficient way. This can be done in two ways
\begin{itemize}
  \item The first method is by speeding up the power spectrum and likelihood
calculations. Interpolation based methods such as \texttt{CMBWarp}
\cite{Jimenez:2004ct}, \texttt{Pico} \cite{Fendt:2006uh} and
\texttt{CosmoNet} \cite{Auld:2006pm} are essentially fast methods of
computing the power spectrum and likelihood from a pre-computed
points on a grid. Also 
\cite{Doran:2005ep} proposed a method to speed up the power spectrum calculation and a similar method is being used in the current version of CAMB. These
methods help MCMC steps to be taken faster.
  \item The second class of methods which can be used in parallel with the first
one improve the efficiency of the MCMC samplers. Therefor shorter
MCMC chains will be needed for a reliable parameter estimation.
Examples of these methods are \emph{COG sampler}
\cite{Slosar:2003da},  \emph{Metropolis sampler with optimal step-size}
\cite{Dunkley:2004sv} and the \emph{slice sampler} in the latest
version of \texttt{CosmoMC}. In this paper, we apply Hamlitonian 
methods to cosmological problems.  As we will show, these methods 
significantly improve the efficency of the MCMC samplers.
\end{itemize}

\emph{Hamiltonian Monte Carlo} (HMC) as first introduced
by \cite{Duan}  is a Markov chain Monte Carlo (MCMC) technique built
upon the basic principle of Hamiltonian mechanics. Its potential in
Bayesian computation makes it a very effective means for exploring
complex posterior distributions. Hamiltonian Monte Carlo is a MCMC
technique in which a momentum variable is introduced for each
parameter of the target probability distribution function (pdf). In
analogy to a physical system, a Hamiltonian $H$ is defined as a
kinetic energy involving the momenta plus a potential energy $U$,
where $U$ is minus the logarithm of the target density. Hamiltonian
dynamics allows one to move along trajectories of constant energy,
taking large jumps in the parameter space with relatively few
evaluations of $U$ and its gradient. The Hamiltonian algorithm
alternates between picking a new momentum vector and following such
trajectories. The efficiency of the Hamiltonian method for
multidimensional isotropic Gaussian pdf's is shown to remain
constant even at high dimensions. Almost all of the traditional 
 MCMC algorithms suffer from the inefficiency caused by the
random walk nature of the Metropolis algorithm. The HMC
algorithms  and some variants on slice sampling 
can be used to prevent the walker from doubling back and
hence to obtain faster convergence and better efficiency.

\section{Bayesian Inference with Markov Chain Monte Carlo}
Bayesian inference in brief is a statistical method by which unknown
properties of a system may be inferred from observation. In Bayesian
inference method we model the system of interest by some model with
parameters $\{x_i\}$. We also introduce a likelihood function to
quantify the probability of the observed data, given the above
model, $f(data|\{x_i\})$.
The Bayesian approach to statistical inference then states that
\begin{equation}
\pi(\{x_i\}|data) \propto f(data|\{x_i\}) p(\{x_i\}).
\end{equation}
 The
distribution $\pi(\{x_i\}|data)$ is referred to as the posterior
distribution and represents our beliefs about the model parameters
after having observed the data. The distribution  $p(\{x_i\})$ is
referred to as the prior and is concerned with our original beliefs
about the model parameters, $\{x_i\}$ before observing the data.
Thus, Bayes' theorem allows us to update our prior beliefs on the
basis of the observed data to obtain our posterior beliefs, which
form the basis for our inference.

Often we are interested in the marginalized distributions of the
parameters
\begin{equation}
\langle \, x_k \,|\, data\, \rangle = \int x_k \, \pi(\{x_i\}|data)
dx_1 dx_2 \cdots dx_D,
\end{equation}
where $D$ is the number of dimensions of the parameter space. At
large $D$, the above integral becomes very hard (and very quickly
impossible) to do. The Monte Carlo principle helps us to carry out
such integrals.

\subsection{Monte Carlo Principle}
The basic idea of Monte Carlo simulation is to draw an independent
and identically-distributed set of samples
$x_i={x_1,x_2,\cdots,x_N}$ from a target density  $p(x)$. The target
density can be approximated by these samples as
\begin{equation}
p_N(x) = \frac{1}{N} \sum_{i=1}^{N}{\delta_{(x,x_i)}}
\end{equation}
where $\delta_{(x,x_i)}$ is the Kronecker delta function.
Consequently, samples from the chain can be used to approximate the
integrals with tractable sums. An integral
\begin{equation}
I = \int{f(x)p(x)dx}
\end{equation}
can be approximated by averaging the function over the chain
\begin{equation}
I_N( f )=\frac{1}{N}\sum_{i=1}^{N} f(x_i).
\end{equation}
Often the posterior mean, $\overline{x}$, and the variance,
$(x-\overline{x})^2$,  are of interest. The can be obtained using
$f(x) = x$ and $f(x) = (x-\overline{x})^2$, respectively.  Drawing
independent samples from $p(x)$ is straight forward when it has a
standard form (for example Gaussian). But in most of the problems,
this is not the case and we have to use more sophisticated
techniques such as Markov Chain Monte Carlo (MCMC). The MCMC technique
has opened up the possibility of applying Bayesian analysis to
complex analysis problems.
\subsection{Markov Chain Monte Carlo Methods}
Markov Chain Monte Carlo is a Markov chain\footnote{A Markov chain
is a series of random variables $x_1,x_2,\cdots,x_N$ such that  the
value at any point in the sequence depends only upon the preceding
value, that is, given a state $x_{i-1}$, the state $x_i$ is
independent of all earlier states except $x_{i-1}$.} that is
constructed to generate samples from a desired target density,
$p(x)$.

The most popular MCMC algorithm is the \emph{Metropolis-Hastings}
algorithm. In this algorithm one takes a trial step, $x^*$ given the
current state, $x$, from an easy-to-sample proposal distribution
$q(x^*|x)$. This trial step is then accepted with the probability of
\begin{equation}
min\{1,\frac{p(x^*)q(x^*|x)}{p(x)q(x|x^*)}\}
\end{equation}
otherwise it remains at $x$. A simple instance of this algorithm is
the \emph{Metropolis algorithm} that uses a symmetric proposal
distribution
\begin{equation}
q(x^*|x) = q(x|x^*)
\end{equation}
and hence the acceptance ratio simplifies to
\begin{equation}
\label{metrop} min\{1,\frac{p(x^*)}{p(x)}\}
\end{equation}
If  $q(x^*|x)=f(\Delta x= x-x^*)$ for some arbitrary function $f$,
then the kernel driving the chain is a random walk and the proposed
step is of the form $x^*= x + \Delta x$, where $\Delta x \sim f$.
There are many common choices for $f$ including the uniform
distribution on the unit disk, multivariate normal or a
$t$-distribution. All of these distributions are symmetric, hence
the acceptance probability has the simple form of eqn.(\ref{metrop})
The algorithm below shows how this is done in practice
\medskip

\newcommand{\Uniform}{\mbox{Uniform}}
\newcommand{\localtt}{\sf}
\begin{center}
\begin{minipage}{.75\textwidth}
\noindent \sf
 \hspace{1.3in}   \textbf{Random Walk Metropolis}
\\{\tt 1:} initialize $x_0$
\\{\tt 2:}  for i = 1 to $N_{steps}$
\\{\tt 3:} \hspace{0.3in} sample $\Delta x$ from the proposal distribution: $\Delta x
\sim q(\Delta x|x)$
\\{\tt 4:}  \hspace{0.3in} $x^* = x + \Delta x$
\\{\tt 5:}  \hspace{0.3in} draw $\alpha  \sim \Uniform(0,1)$
\\{\tt 6:} \hspace{0.3in} if $\alpha < \min\{1,\frac{p(x^*)}{p(x)}\}$
\\{\tt 7:} \hspace{0.7in}$x_i = x^*$
\\{\tt 8:} \hspace{0.3in} else
\\{\tt 9:} \hspace{0.7in}$x_i = x_{(i-1)}$
\\{\tt 10:} end for
\\
\end{minipage}
\end{center}
\medskip

In order for this method to be effective, samples must be as
uncorrelated and independent as possible. A common problem of MCMC
techniques is that samples can be highly correlated. This makes the
method inefficient. In this case although the samples are drawn from
the correct distribution, they sample the target density very
slowly. And a huge number of samples might be needed for  reliable
estimates.

\subsection{Hamiltonian Monte Carlo}
Hamiltonian Monte Carlo (HMC) belongs to the class of MCMC
algorithms with auxiliary variable samplers\footnote{Another method
of this kind is \emph{slice sampling} that has been exploited in
CosmoMC package.}. HMC is a MCMC technique in which a momentum
variable is introduced for each parameter of the target probability
distribution function (pdf). In analogy to a physical system, a
Hamiltonian $H$ is defined as a kinetic energy involving the momenta
plus a potential energy $U$, where $U$ is minus the logarithm of the
target density. Hamiltonian dynamics allows one to move along
trajectories of constant energy, taking large jumps in the parameter
space with relatively few evaluations of $U$ and its gradient. The
Hamiltonian algorithm alternates between picking a new momentum
vector and following such trajectories. This algorithm is designed
to improve mixing in high dimensions by getting more uncorrelated
and independent samples. To explain this in more detail, suppose we
wish to sample from the distribution $p(\mathbf{x})$, where
$\mathbf{x}\in \mathcal{R}^D$ ($\mathcal{R}^D$ is the
$D$-dimensional parameter space of our problem). We augment each
parameter $x_i$ with an auxiliary conjugate momentum $u_i$, and
define the potential energy
\begin{equation}\label{U}
    U(\mathbf{x}) = - \ln{p(\mathbf{x})},
\end{equation}
and the Hamiltonian
\begin{equation}\label{H}
    H(\mathbf{x},\mathbf{u}) = U(\mathbf{x}) + K(\mathbf{u}),
\end{equation}
where $K(\mathbf{u})= \mathbf{u}^T \mathbf{u}/2$ is the kinetic
energy. This is used to create samples from the extended target
density
\begin{eqnarray}
p(\mathbf{x},\mathbf{u}) &\propto& \exp(-H(\mathbf{x},\mathbf{u}))
\\ \nonumber
&=& \exp(-U(\mathbf{x})) \exp(-K(\mathbf{u}))
\\ \nonumber
&=& p(\mathbf{x}) \mathcal{N}(\mathbf{u};0,1),
\end{eqnarray}
where $\mathcal{N}(\mathbf{u};0,1)$ is the $D$-dimensional normal
distribution with zero mean and unit variance. This density is
separable, so the marginal distribution of $\mathbf{x}$ is the
desired distribution $ p(\mathbf{x}) $. This means if we can sample
$(\mathbf{x},\mathbf{u})$ from the extended distribution
$p(\mathbf{x},\mathbf{u})$, then the marginal distribution of
$\mathbf{x}$ is exactly the target distribution $ p(\mathbf{x}) $.

Each step in HMC consists of drawing a new pair of samples
$(\mathbf{x},\mathbf{u})$ according to $p(\mathbf{x},\mathbf{u})$
starting from the current value of $\mathbf{x}$ and generating a
Gaussian random variable $\mathbf{u}$. These are our initial
conditions. The time evolution of this system is then governed by
Hamiltonian equations of motion
\begin{eqnarray}
\dot{x}_i&=&u_i
\\ \nonumber
\dot{u}_i&= &-\frac{\partial H}{\partial x_i}
\end{eqnarray}
Because the Hamiltonian dynamics is time-reversible,
volume-preserving and total energy preserving, if the dynamics are
simulated exactly, the resulting moves will leave the extended
target density, $p(\mathbf{x},\mathbf{u})$, invariant. That is, if
we start from $(\mathbf{x}_{(0)},\mathbf{u}_{(0)})\sim p$, then
after the system evolves for time $t$, the new configuration at time
$t$, $(\mathbf{x}_{(t)},\mathbf{u}_{(t)})$ also follows the
distribution $p$. But in practice the dynamics are simulated with a
finite step-size, and as a result, $H$ won't remain invariant. The
Hamiltonian dynamics in practice is simulated by the \emph{leapfrog
algorithm} with a small step-size $\epsilon$
\begin{eqnarray}
u_i(t+{\epsilon\over{2}})&=&u_i(t)-{\epsilon\over{2}}\left(\frac{\partial
U}{\partial x_i}\right)_{\mathbf{x}_{(t)}}
\\ \nonumber
 x_i(t+\epsilon)&=& x_i(t) + \epsilon u_i(t+{\epsilon\over{2}})
\\ \nonumber
u_i(t+{\epsilon\over{2}})&=&u_i(t)-{\epsilon\over{2}}\left(\frac{\partial
U}{\partial x_i}\right)_{\mathbf{x}_{(t+\epsilon)}}.
\end{eqnarray}
Each leapfrog move is volume-preserving and time-reversible. But it
does not keep the $H$ constant. The effect of inexact simulation
introduced by the non-zero step-size can be eliminated by the
Metropolis rule: the point reached by following the dynamics is
accepted with the probability of
\begin{equation}
\min\{1,e^{-(H(x^*,u^*)-H(x,u))}\}.
\end{equation}
The algorithm of HMC is shown below
\medskip
\begin{center}
\begin{minipage}{.75\textwidth}
\noindent \sf
 \hspace{1.3in}   \textbf{Hamiltonian Monte Carlo}
\\{\tt 1:} initialize $\mathbf{x}_{(0)}$
\\{\tt 2:}  for i = 1 to $N_{samples}$
\\{\tt 3:} \hspace{0.3in} $\mathbf{u} \sim \mathcal{N}(0,1)$
\\{\tt 4:} \hspace{0.3in} $(\mathbf{x}^*_{(0)},\mathbf{u}^*_{(0)})= (\mathbf{x}_{(i-1)
},\mathbf{u})$
\\{\tt 5:}  \hspace{0.3in} for j = 1 to N
\\{\tt 6:}  \hspace{0.7in} make a leapfrog move: $(\mathbf{x}^*_{(j-1)
},\mathbf{u}^*_{(j-1) }) \rightarrow (\mathbf{x}^*_{(j)
},\mathbf{u}^*_{(j) })$
\\{\tt 7:}  \hspace{0.3in} end for
\\{\tt 8:}  \hspace{0.3in} $(\mathbf{x}^*,\mathbf{u}^*)= (\mathbf{x}_{(N)
},\mathbf{u}_{(N) })$
\\{\tt 9:}  \hspace{0.3in} draw $\alpha  \sim \Uniform(0,1)$
\\{\tt 10:} \hspace{0.3in} if $\alpha < \min\{1,e^{-(H(\mathbf{x}^*,\mathbf{u}^*)-H(\mathbf{x},\mathbf{u}))}\}$
\\{\tt 11:} \hspace{0.7in} $\mathbf{x}_{(i)}=\mathbf{x}^*$
\\{\tt 12:} \hspace{0.3in} else
\\{\tt 13:}
\hspace{0.7in}$\mathbf{x}_{(i)}=\mathbf{x}_{(i-1)}$
\\{\tt 14:} end for
\\
\end{minipage}
\end{center}
\medskip

To see how this method works, let's consider the simplest target
distribution: a one dimensional Gaussian distribution. The
Hamiltonian for this case is that of a harmonic oscillator
\begin{equation} \label{harmonicoscillator}
H(x,u) = -\frac{x^2}{2\sigma^2} + \frac{u^2}{2}.
\end{equation}
The phase space trajectories of this system are ellipses with their
size(area) proportional to $H$. In practice, a HMC algorithm would
run along such an ellipse while performing leapfrog moves. If the
algorithm consisted of only this leapfrog update, it would not be
irreducible. Because all points generated from a starting point will
never leave an ellipse of constant $H$ and hence will not visit all
points in the parameter space. The Gibbs update of the momentum
(line {\sf{3}} of the HMC algorithm shown above) rectifies this
situation. At the beginning of every iteration the momentum is
updated in such a way that that the Markov chain gets a chance of
reaching all other values of $H$ and hence visiting every point in
the parameter space after enough number of iterations.

In most practical cases (and almost always in astrophysics), we
don't know the analytical form of the target distribution,
$p(\mathbf{x})$. Therefor one can not derive the gradient of the
logarithm of the $p(\mathbf{x})$ in a closed form. Usually computing
the $p(\mathbf{x})$ is expensive enough that makes it impossible to
compute the gradient numerically either. In this case, one performs
an exploratory run of a simple MCMC method to get an idea of the
$p(\mathbf{x})$. This exploratory run can be used to fit a function
 to the minus logarithm of target density, $U(\mathbf{x})$. This is
like estimating the proposal distribution in traditional MCMC
algorithms except that we are no more limited to ``easy-to-sample''
distributions.  Gradients of this proposal distribution can be used
as estimates of gradients of the actual likelihood. This is
explained in a cosmological context in Section \ref{CosmoHMC}.

There are two parameters to tune while working with the HMC: 
step-size $\epsilon$ and the number of leapfrog steps to be taken.
As it was mentioned above, $\epsilon$ should be chosen small enough
to do the Hamiltonian dynamics correctly. Big step-sizes 
will bring errors into the simulation and won't keep the 
total energy constant. Hence the acceptance rate decreases.
In the simplest 
case, \emph{e.g.} eqn. \ref{harmonicoscillator}, it can be 
immediately seen that a reasonable   $\epsilon$ should be smaller than 
the ``natural'' time-scale of the system which is
\be
\epsilon < (\frac{\partial^2 H}{\partial^2 x})^{-1/2}.
\ee
The number of leapfrog steps, $N$, should be large enough to take the walker 
far from the starting point. A choice of $N\epsilon=\sigma_0$ has worked well
 for most of the problems we deal with. It helps if the $N\epsilon$ 
is randomized in order to avoid the resonance condition \cite{neal}, and
it also improves the efficiency. However the choice of these two parameters 
may vary from a problem to another. The best choice would be obtained by 
monitoring the acceptance rate and the efficiency of the resulting chains.

The HMC resolves some inefficiencies of the traditional MCMC methods
such as low acceptance rate, high correlations between the
consequent steps, slow mixing and slow convergence. To be able to
compare different methods, we need a set of reliable diagnostic
tools. The next section introduces some of these tools and they will
be used in the rest of this paper to compare different techniques. For
more details on the HMC technique see references \cite{Brooks98},
\cite{Chen2001}, \cite{neal}, \cite{ChooThesis}, \cite{McKay},
\cite{Rasmussen}, \cite{Andrieu} and \cite{Hanson}.
\section{Convergence Diagnostics and Efficiency}
While working with MCMC algorithms there are a number 
of important questions that one should answer. These questions are
Are we sampling from the distribution of interest? 
 Are we are seeing the full distribution?  
Are we able to estimate parameters of interest to a desired precision?  
And are we doing all this efficiently?  
In order to answer the above questions,
 several diagnostic tests have been developed so far. 
In this paper we will use four methods to quantify the convergence and 
efficiency of MCMC algorithms. \\
\textbf{Autocorrelation:} To examine relationships of successive
steps of a chain, $x_i$,the autocorrelation function (ACF) can be
used. The autocorrelation function at lag $l$ is
\begin{eqnarray} \label{acorr}
\rho(l)&=&\frac{ {\rm Cov}({x}_i, {x}_{i+l}) } { {\rm Var}({x}_i) }
\\ \nonumber
&=&\frac{\sum_{i=1}^{n-l} ({x}_i- {\overline {x}})
({x}_{i+l}-{\overline {x}}) } {\sum_{i=1}^n ({x}_i-{\overline
{x}})^2}
\end{eqnarray}
An ideal sampler with no correlations between successive elements of
the chain will have an autocorrelation that quickly vanishes. High
autocorrelations within chains indicate slow mixing and, usually,
slow convergence. It could be useful to thin out a chain with high
autocorrelations before calculating summary statistics: a thinned
chain may contain most of the information, but take up less
space in memory. \\ 
\textbf{Autocorrelation Length:} The autocorrelation length of a
chain is defined as
\begin{equation} \label{acorrtime}
L = 1 + 2 \sum_{l=1}^{l_{max}} \rho(l)
\end{equation}
Where $l_{max}$ is the maximum lag. It is good to stop summing at a
reasonable $l_{max}$ beyond which the autocorrelation becomes noisy
to avoid introducing errors into the autocorrelation length
calculation. An ideal sampler will have $L = 1$, and larger $L$
would indicate more correlation between the data.\\
\textbf{Power Spectrum:} The power spectrum of an infinite chain is
defined as
\begin{equation} \label{pk}
\langle X(k) X^*(k') \rangle = \delta(k-k') P(k),
\end{equation}
where $X(k)$ is the Fourier transform of the chain, $x_i$.
 In practice, the power spectrum is estimated from a finite chain
in the following way
\begin{equation}
P(k) = |\alpha_k|^2,
\end{equation}
in which $\alpha_k$ are the discrete (inverse) Fourier transform of
the chain divided by the square root of the number of samples in the
chain, $N$. And $k=2\pi j/N$ for $j=1,\cdots,N/2-1$. An ideal
sampler will have a flat power spectrum. While actual MCMC chains
have correlations on small scales and therefore their power spectrum
bends on small scales \cite{Dunkley:2004sv}. The $P(k)$ at $k=0$ is
an estimate of the sample mean variance, $\sigma_{\bar{x}}^2(N)$,
\begin{equation}\label{P0}
\sigma_{\bar{x}}^2(N) = \frac{P_0}{N}, \,\,\,\,\,\,\,\,\,\, P_0 =
P(k=0).
\end{equation}
\\
 \textbf{Efficiency:} The statistical
efficiency of an MCMC sequence is defined as the ratio of the number
of independent draws from the target pdf to the number of MCMC
iterations required to achieve the same variance in an estimated
quantity. The efficiency is defined as
\begin{equation} \label{E}
E = \lim_{N\rightarrow \infty}
\frac{\sigma_0^2/N}{\sigma_{\bar{x}}^2(N)}
\end{equation}
where $\sigma_0$ is the variance of the underlying distribution and
$\sigma_{\bar{x}}$ is the variance of the sample mean from the
chain. $E^{-1}$ is an estimate of the factor by which the MCMC chain
is longer than an ideal chain. Comparing eqn.(\ref{E}) with
eqn.(\ref{P0}), we see that the efficiency of an MCMC chain is given
by $P_0$
\begin{equation} \label{EvsP0}
E = \frac{\sigma_0^2}{P_0}
\end{equation}
\begin{figure}[]
  \includegraphics[]{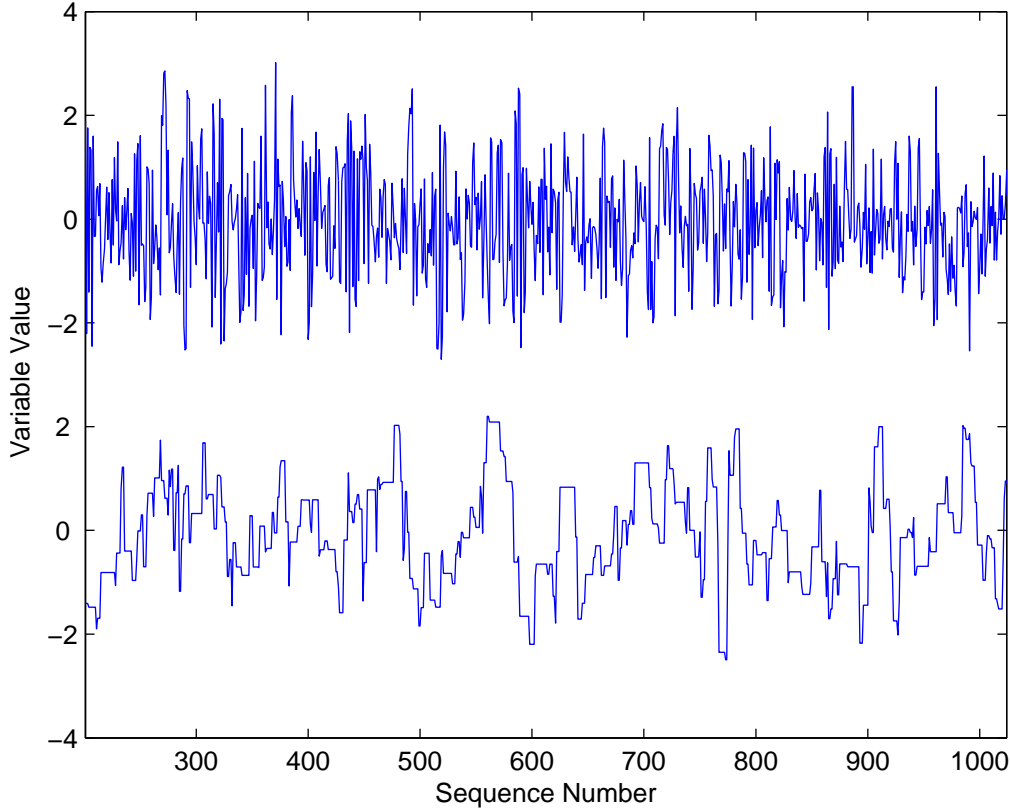}\\
  \caption{Samples drawn from an isotropic six-dimensional Gaussian distribution using the HMC (top) and the Metropolis algorithm with optimal step-size (bottom). }\label{Chains_comparison}
\end{figure}
For more diagnostics see references \cite{MCMCstuff}, \cite{Cowles}
and references there in.
\section{Applications of HMC }

We start from the simplest distribution, $D$-dimensional Gaussian
distribution
\begin{equation}
E(\mathbf{x}) = \frac{1}{2}\mathbf{x}^TC^{-1}\mathbf{x},
\,\,\,\,\,\,\,\,\,\,\,\,\,\,\,\, C_{ij} = \sigma^2_0 \delta_{ij}
\end{equation}
with the gradient
\begin{equation} \Delta E(\mathbf{x}) =
C^{-1}\mathbf{x}
\end{equation}
A 6-dimensional Gaussian distribution (as given above) with
$\sigma_0=1$ is sampled with a chain of length $N_{samples}=8192$
using two methods: the Hamiltonian Monte Carlo method with $N=100$
leapfrog steps taken at each iteration at a step-size chosen such
that $N \epsilon = \sigma_0$,  and a Metropolis algorithm
that uses a Gaussian proposal distribution with a step-size
$\sigma_T/\sigma_0 \approx 2.4 / \sqrt{D}$.

\begin{figure}[]
  \includegraphics{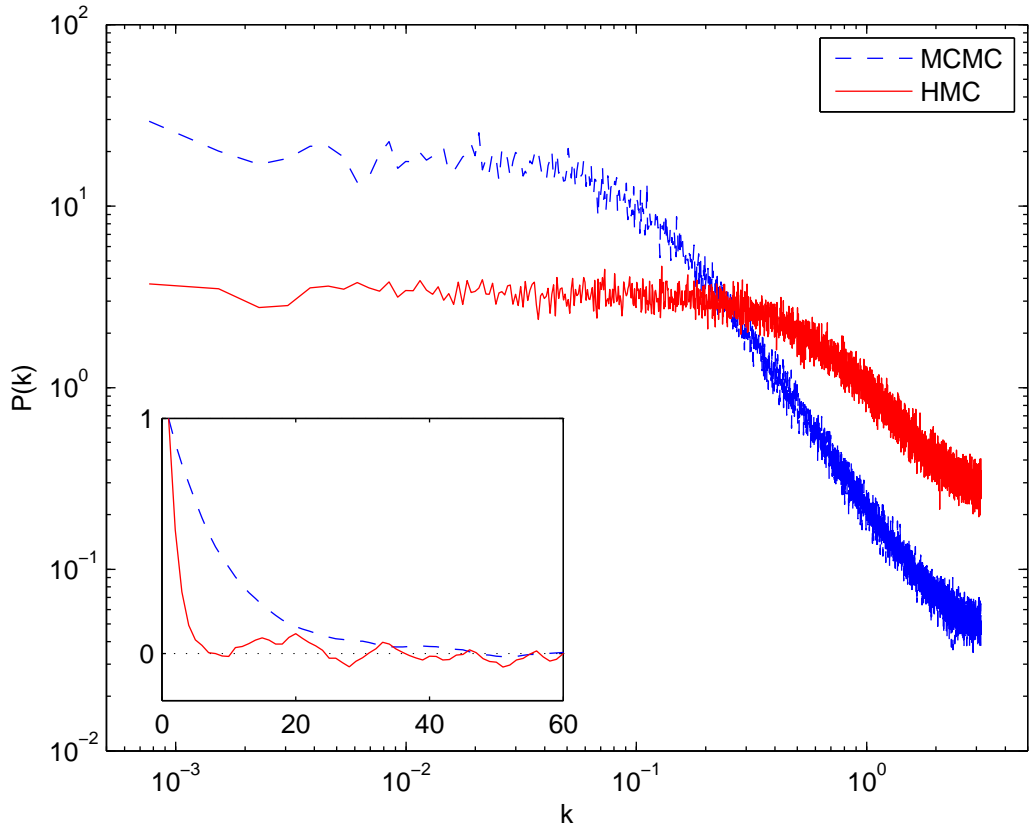}\\
  \caption{Power spectrum, $P(k)$, for the Hamiltonian Monte Carlo (solid red line)
 compared to that of a traditional MCMC method with an optimal step-size (dashed blue line).}\label{Pkavg_6D}
\end{figure}

Fig. \ref{Chains_comparison} shows the chains generated by the two
algorithms. These chains are used to compare the two methods. The
acceptance rate for the HMC is $99\%$ while it is $25\%$ for the
MCMC.  HMC improves the mixing by accepting independent samples $4$
times more than the traditional MCMC chain. Correlations between the
successive steps in these methods are shown in Fig. \ref{Pkavg_6D}.
The small panel shows the autocorrelations: the HMC autocorrelation
dies off much faster than that of the MCMC. This feature is seen
much better in the power spectrum of the chains. The HMC takes a
smaller time to enter the white-noise regime at large scales than
MCMC. This is given by the scale on which the power spectrum turns
over and becomes flat which is around $k=0.1$ for the  HMC and
$k=0.02$ for the MCMC.
The efficiency of the chains can be read directly from the power
spectra of Fig. \ref{Pkavg_6D}. The efficiency is given by
$\sigma_0/P_0$ and since $\sigma_0=1$, it is simply equal to
$1/P_0$. The ratio of efficiencies are
\begin{eqnarray}
\frac{E_{HMC}}{E_{MCMC}}=\frac{P_0^{MCMC}}{P_0^{HMC}} \approx 6.
\end{eqnarray}
Therefore in 6 dimensions an MCMC chain is 6 times longer than an
HMC chain yielding the same performance. Later we will show that
this number scales as the number of dimensions and hence the HMC
does better than the MCMC on higher dimensions.
\begin{figure}[]
  \includegraphics{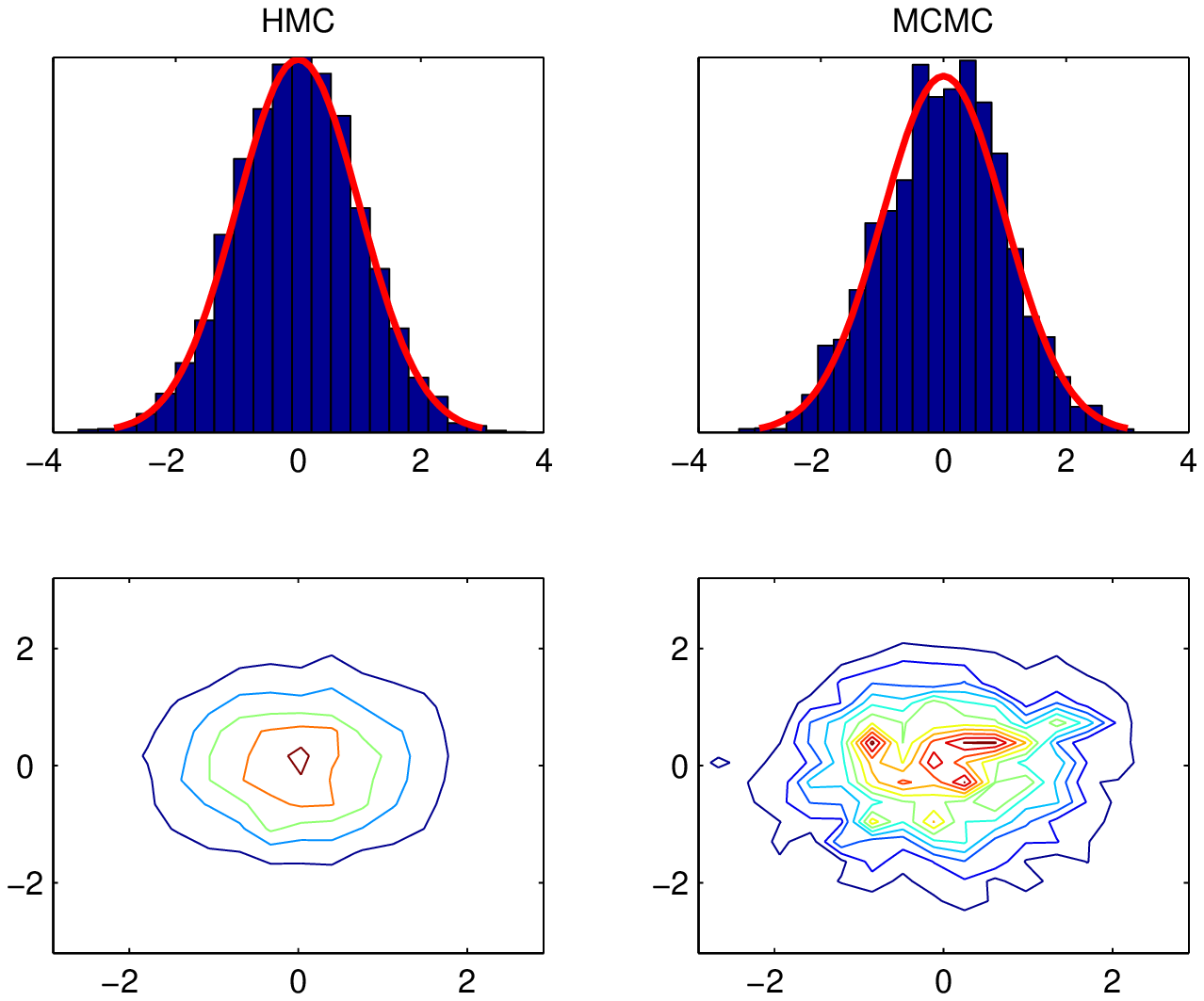}\\
  \caption{Marginalized distributions and two-dimensional contour plots of
the 6-dimensional Gaussian model sampled with chains of length
$N_{samples}=8192$ using the HMC (\emph{left}) and  MCMC
(\emph{right}). }\label{MCMCvsHMC_ContourandHist}
\end{figure}

This can be seen more clearly in Fig.
\ref{MCMCvsHMC_ContourandHist}, where histograms of marginal
distributions and two-dimensional contour plots are plotted for the
both chains. For the same length of chains, the HMC gives more
accurate results.

\subsection{Efficiency of the HMC }
\begin{figure}
  \includegraphics[width=0.6\textwidth]{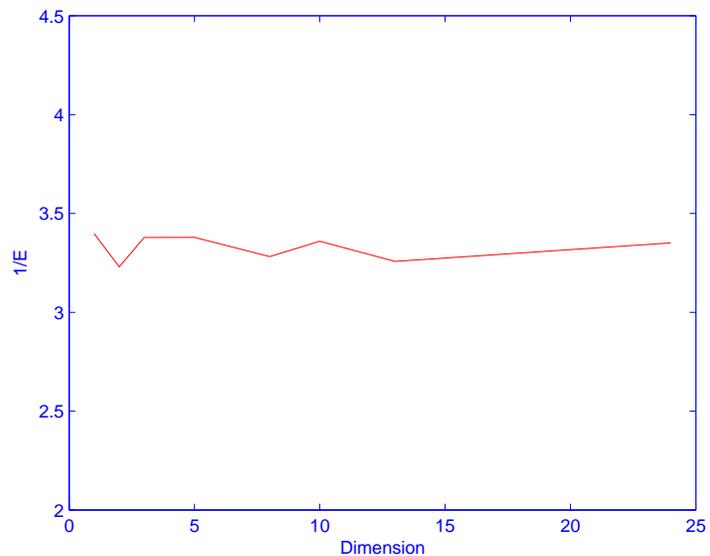}\\
  \caption{Average efficiency of the HMC vs. the number of dimensions. }\label{efficiencyvsd}
\end{figure}

  The exercise above can be repeated
for other dimensions to see how the efficiency of the HMC chains are
scales with the number of dimensions of the problem of interest.
This can be done in two ways
\begin{enumerate}
  \item by comparing the sample mean variance to an ideal chain;
eqn.(\ref{E}),
  \item by computing the power spectrum and deriving the $P(k=0)$;
eqn.(\ref{EvsP0}).
\end{enumerate}
We computed the efficiency of the HMC using the first method and
cross checked it with the second method to confirm our results. The
analysis are done for the HMC algorithm with $N\epsilon=1$. The
efficiency is plotted versus the number of dimensions in Figure
\ref{efficiencyvsd} and shows that it remains constant even in high
dimensions.

As a comparison, the optimal efficiency of the MCMC method is
inversely proportional to the number of dimensions; $E \approx
{1\over 3.3 D}$ \cite{Dunkley:2004sv}. And hence the ratio 
of the two efficiencies is
proportional to the number of dimensions
\begin{eqnarray}
\frac{E_{HMC}}{E_{MCMC}} \approx D.
\end{eqnarray}
Therefore in $D$-dimensions an HMC chain is $D$ times shorter than
an MCMC chain yielding the same performance.
\subsection{Non-Gaussian And Curved Distributions }
In 2 or more dimensions, the distribution of our interest may be very 
complicated. Sampling from non-Gaussian or curved distributions can be difficult and inefficient. In these cases usually a re-parametrization of the problem 
helps a lot to transform the distribution to a relatively simpler one which is 
easier to sample from \cite{Verde:2003ey}. 
\begin{figure}[]
  \includegraphics[width=\textwidth]{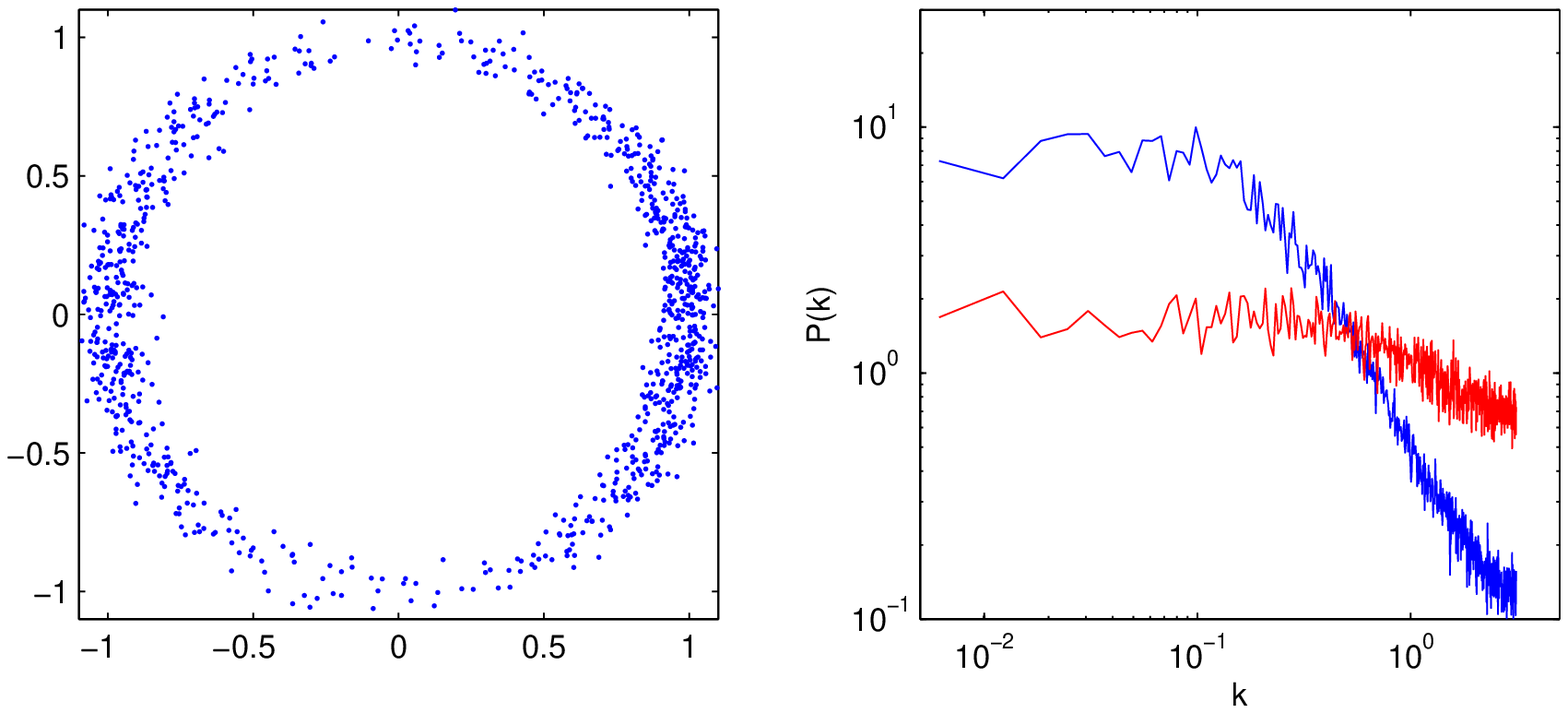}\\
  \caption{\emph{left: } The first 1000 samples drawn from the 
crescent-like distribution of eqn. \ref{crescent_formula} and their power spectra, $P(k)$ (\emph{right}).}\label{crescent}
\end{figure}
The HMC can improve the efficiency of sampling from these distributions. 
To demonstrate this, we choose the worst case scenario of \cite{Dunkley:2004sv}; a thin, curved, non-Gaussian distribution given by
\be
\label{crescent_formula}
E = \frac{(x^2+y^2-1)^2}{8\sigma_1^2} + \frac{y^2}{2\sigma_2^2}.
\ee
Sampling the above distribution in the present form is very inefficient with 
the traditional MCMC methods  \cite{Dunkley:2004sv}. In contrast, the HMC 
can easily sample that.  Figure \ref{crescent} shows the first 1000 samples drawn from the above distribution and their power spectra. The widths are chosen such that the dimensionless parameter $\sigma_2^2/\sigma_1 =5$. As it can be seen in the figure, the chains have reached the equillibrium shortly and are sampling from the desired distribution. This is an idealized case where we exactly know the gradient at every point in the parameter space. However in a real-world problem we don't know the gradients very accurately. We will discuss this in the next section and by applying this method to cosmological parameter estimation we will show that even a simple guess of the gradients will work. But better guesses will result in more efficient estimations.

\section{Cosmological Parameter Estimation with Hamiltonian Monte Carlo}\label{CosmoHMC}
The HMC is shown to be very promising in reducing the length of
Markov chains for a reliable inference. Hence, this method can
potentially speed up the cosmological parameter estimation by
reducing the number of samples of a Markov chain. Taking a closer
look at a typical algorithm for parameter estimation we will see the
bottle necks of these algorithms are
\begin{enumerate}
  \item Computing the power spectrum: at each step the CMB power
spectrum should be computed for the new set of parameters. As the
new experiments push to the larger $l$, computing the power spectrum
becomes more time consuming.
  \item Computing the likelihood: at each step, given the power
spectrum, the likelihood should be computed. This is not as slow as
the power spectrum computation, but becomes a bottle neck in long
chains.
  \item Number of samples needed in Markov chains: usually long
Markov chains must be generated to achieve reliable results. Two
practical issues of the MCMC chains are high correlations and low
efficiency that result in long burn in times and high rejection
rate. Specially in high dimensions, very lengthy Markov chains are
needed (the `curse of dimensionality').
\end{enumerate}
The first two problems have attracted a lot of attention recently
and fast methods of computing the power spectrum and likelihood have
been designed \cite{Jimenez:2004ct}, \cite{Fendt:2006uh} and
\cite{Auld:2006pm}. All of these methods try to speed up the
parameter estimation by making each step taken in a shorter time,
but due to the nature of the traditional MCMC methods, still long
chains are needed to obtain reliable results. An orthogonal and
complimentary approach is to reduce the length of the Markov chains.
This can be done by the HMC method.  HMC can be done in ordinary
cosmological parameter estimation codes. The method is simple and it
is straightforward to add a module to a standard cosmological
parameter estimation package, like CosmoMC, to perform that. The HMC
algorithm needs to compute the gradient of the minus logarithm of
the likelihood in each step. This can be done using an auxiliary
distribution that mimics the minus logarithm of the likelihood
function but is fast to compute. The leapfrogs can then be taken
with no major additional cost of time.
\subsection{Computing The Gradients: Zeroth Order Approximation }
The simplest choice for an auxiliary function to estimate the
gradients of the minus logarithm of the likelihood is
\begin{equation}
\frac{\mathcal{L} - \overline{\mathcal{L}}}{\sigma_{\mathcal{L}}} =
({\bf x} - \overline{{\bf x}})^\dagger C^{-1} ({\bf x} -
\overline{{\bf x}}),
\end{equation}
and the gradient would be
\begin{eqnarray} \label{gradient_gaussian}
\frac{\partial \mathcal{L}}{\partial x_i} &=& \sigma_{ \mathcal{L}}
\left( \tilde{C}_{ij}^{-1} ( x_j - \overline{ x}_j) + ( x_j -
\overline{ x}_j) \tilde{C}_{ji}^{-1}\right)
\\ \nonumber
&=&\, 2\,\sigma_{ \mathcal{L}}\, \tilde{C}_{ij}^{-1} \,( x_j -
\overline{ x}_j).
\end{eqnarray}
As an example we sample from the 6-parameter flat LCDM model using
the HMC. An exploratory run of traditional MCMC can be performed to
obtain the fitting parameters $\sigma_{ \mathcal{L}},
\overline{\mathcal{L}}$ and $\overline{\mathbf{x}}$. Using the
simple MCMC option of CosmoMC (\texttt{sampling\_method=1} option is
to use the default Metropolis algorithm in the optimal fast/slow
subspaces) at the exploratory phase  the following fit to the minus
logarithm of the likelihood is found
 \begin{eqnarray}
 \mathcal{L} &\simeq&  ({\bf x} - \overline{{\bf
x}})^\dagger C^{-1} ({\bf x} - \overline{{\bf x}})
\sigma_{\mathcal{L}}  + \overline{\mathcal{L}}
\\ \nonumber
&=& \, 0.5164 \, ({\bf x} - \overline{{\bf x}})^\dagger \, C^{-1} \,
({\bf x} - \overline{{\bf x}}) \, + \,  5626,
\end{eqnarray}
where $C_{ij}$ is the covariance matrix and $\overline{\mathbf{x}}$
 are the means, all estimated from the exploratory chains.
These are all we need, the gradient can then be estimated using eqn.
(\ref{gradient_gaussian}). Running modified CosmoMC with the above
sampling method, yields $81\%$ acceptance rate.  Proposed steps are
rejected $19\%$ of times because of the imperfect estimate of the
gradient of the field which causes inaccurate simulation of
Hamiltonian dynamics. Even this simplest choice of the auxiliary
function improves the acceptance rate by more than a factor of $3$
and reduces the correlations which in turn boosts up the efficiency
of the chain and reduces the burn in time. However, this can be
further improved to achieve a close-to-ideal sampler.
\subsection{Computing The Gradients: Using Lico}
\begin{figure}
  \includegraphics[width=0.6\textwidth]{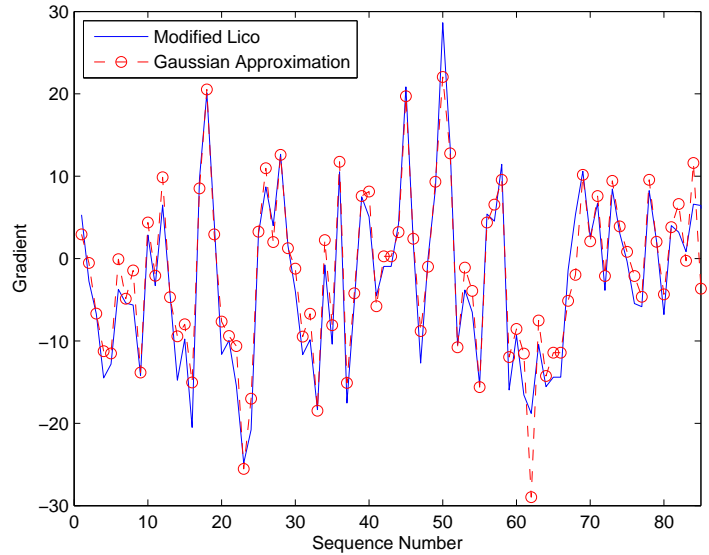}\\
  \caption{Gradients of the minus logarithm of likelihood computed at 85 random positions using a Gaussian approximation to the likelihood and modified Lico. }\label{grads}
\end{figure}
A good feature of the HMC is that we are not limited to Gaussian
auxiliary functions to estimate the gradients from. Hence we can
choose functions that better resemble the minus log likelihood
function. For example a polynomial fit can be made in the following
form
\begin{eqnarray}
\frac{\mathcal{L} - \overline{\mathcal{L}}}{\sigma_{\mathcal{L}}}
&=& c_0 + c_1(\frac{ x_1 -  \overline{x}_1}{\sigma_1}) + c_2(\frac{
x_2 -  \overline{x}_2}{\sigma_2}) + c_3(  \frac{x_1 -
\overline{x}_1}{\sigma_1})(\frac{ x_1 -\overline{x}_1} {\sigma_1} )
\\ \nonumber
&+&  c_4 (  \frac{x_1 - \overline{x}_1}{\sigma_1}) (\frac{ x_2
-\overline{x}_2}{\sigma_2} )+ c_5 ( \frac{ x_2 -
\overline{x}_2}{\sigma_2})(\frac{ x_2 - \overline{x}_2}{\sigma_2} )+
\cdots.
\end{eqnarray}
The coefficients of the above equation are found by \emph{Lico}, the
likelihood routine of the Pico package \cite{Fendt:2006uh}. Lico breaks
the parameter space into 30 regions and performs a fourth order
polynomial fit to the $\mathcal{L}$ in each region separately. It is
possible to modify Lico to compute the gradient of the above
equation. This method must give a much better estimate of the
gradients comparing to the simplest one.
\begin{figure}
  \includegraphics[]{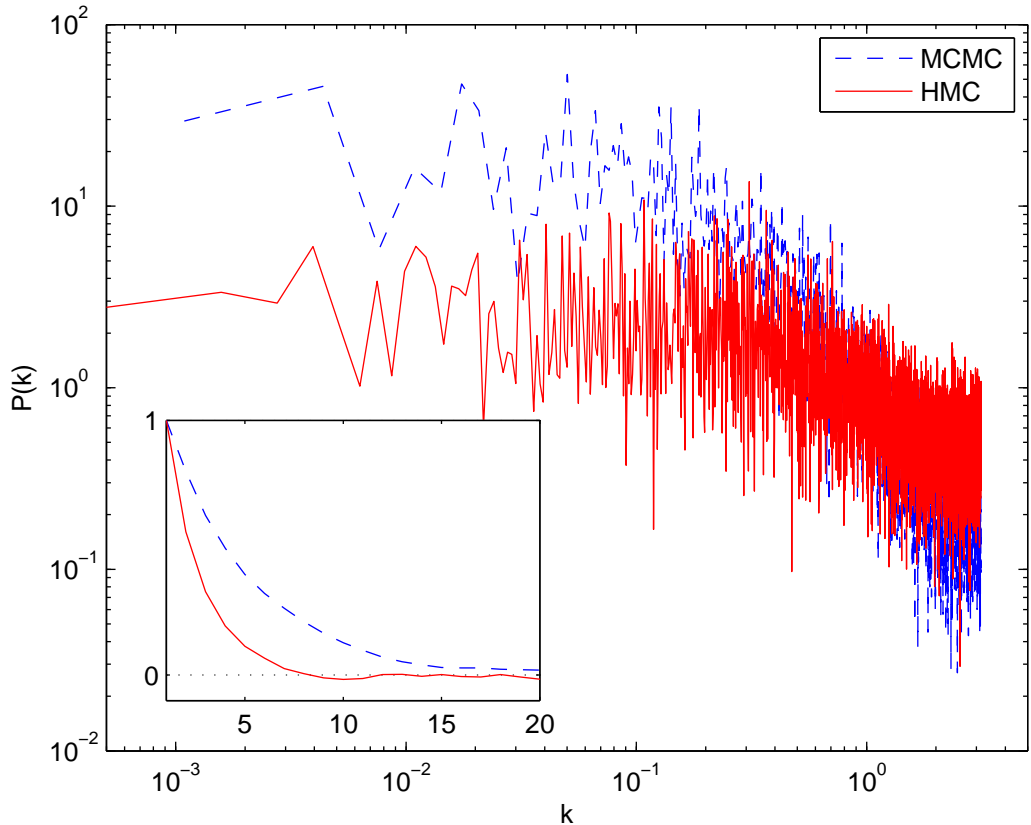}\\
  \caption{Binned power spectra, $P(k)$, of the Markov chains sampled from the 6-parameter flat LCDM  model using the  default Metropolis algorithm of
           \texttt{CosmoMC} (dashed blue line) and the HMC (solid red
            line). The ratio of efficiencies are given by the inverse ratio of $P(0)$ for each method.  The HMC is almost a factor of $6$ more
efficient than the default Metropolis algorithm of \texttt{CosmoMC}. Autocorrelation functions for $\Omega_b h^2$ chains are also plotted in the small panel for comparison. }\label{Pkavg_HMCvsCosmoMC}
\end{figure}

Fig. \ref{grads} shows the differences between the gradients
computed using the two methods. Comparison is done at 100 random
points in the parameter space.

To use this method in the HMC, we have added a module to CosmoMC
that allows CosmoMC to use the above method based on Lico to compute
the gradient of $\mathcal{L}$ whenever the parameters are within the
range over which Pico's regression coefficients are defined. For
parameters outside this range, CosmoMC will continue to use the
Gaussian approximation to compute the gradient.

\subsection{A Comparison between the HMC and the Monte Carlo method of CosmoMC}
\begin{table}
  \centering
\begin{tabular}{|c||c|c|c|c|c|c|| c|}
    \hline
  &   $\Omega_b h^2$  &  $\Omega_c h^2$ &   $\theta$  & $\tau$  &$n_s$& $A_s$ & Average \\
    \hline \hline
MCMC & 11.5& 19.6& 11.7& 28.9& 17.8& 13.5& 17.1\\
HMC &3.2&  2.8&  4.0&  2.9&  3.4&  3.8 & 3.3\\
    \hline
\end{tabular}
  \caption{Autocorrelation lengths of the sampled chains from the 6-parameter flat LCDM model using the HMC and the Metropolis method of
\texttt{CosmoMC}. }\label{acorrtimes}
\end{table}
Having the HMC equipped with the above method of gradient
estimation, we sample the 6-parameter flat LCDM model with an HMC
sampler.  The leapfrog step-sizes are the chosen to be $\epsilon =
0.01\sigma_i$, where $\sigma_i=\sqrt{C_{ii}}$. We randomly draw the
number of leapfrog steps taken at each iteration from a uniform
distribution such that $N\,\epsilon \sim Uniform(0,3) \sigma_i$. We run the
chain to generate $16000$ samples. This time we get $98\%$
acceptance rate, and very low correlations in the chain. We also run
\texttt{CosmoMC} using the simple Monte Carlo option (with the
optimal step-size $\sigma_T = 2.4 \sigma_i/\sqrt{D}$) to generate
the same number of samples. The acceptance rate is $35\%$. The
autocorrelation lengths of the HMC chain for the 6-parameter flat
LCDM model are compared to those of the  Metropolis  method of
\texttt{CosmoMC} in Table \ref{acorrtimes}. The average
autocorrelation length in the HMC chains are smaller than those of
MCMC chains by a factor of $\sim 6$. This can be seen better in the
power spectrum plots of Fig. \ref{Pkavg_HMCvsCosmoMC}.
Autocorrelation functions are also plotted in the small panel for
comparison. The efficiency of the HMC is almost an order of
magnitude better than that of the default Metropolis method of
\texttt{CosmoMC}. Therefore, using the HMC, one needs a much shorter
chain to obtain the same accuracy as the Metropolis method.

\begin{figure}[h]
  \includegraphics[]{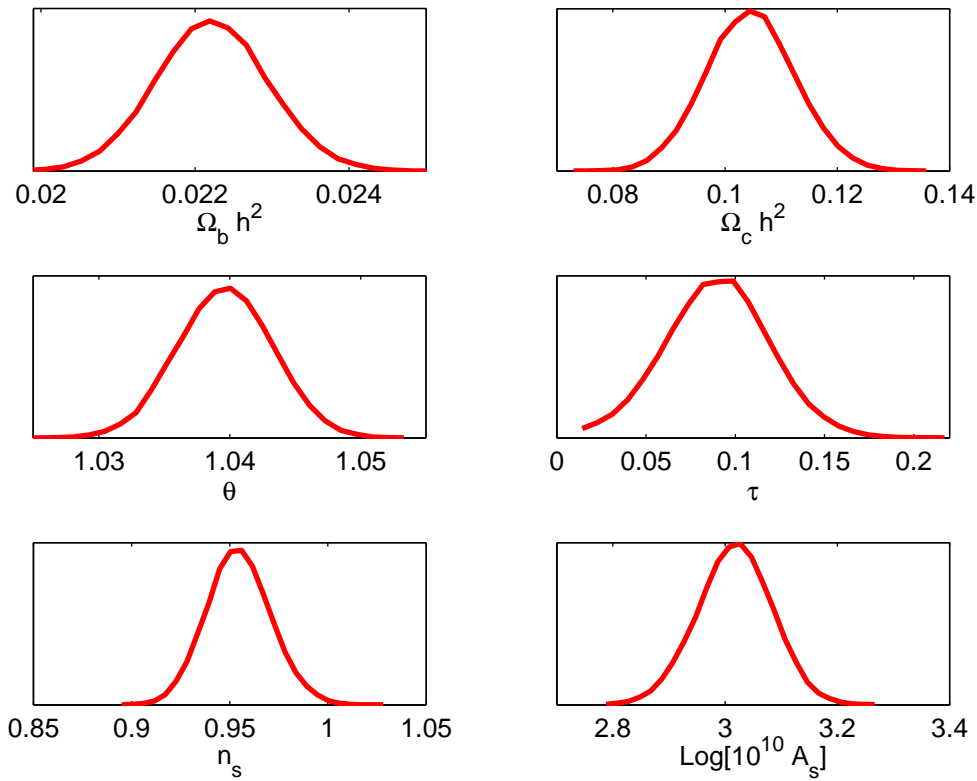}\\
  \caption{Marginalized distributions of the 6-parameter
flat LCDM model obtained from the HMC.}\label{}
\end{figure}

\section{Discussion and Conclusions}
With the future data of the upcoming experiments like ACT \cite{ACT}
and Planck \cite{Plank} that will map the CMB sky on small scales
(large $l$), the need to efficient and accurate methods of parameter
estimation will be more. Besides finding methods of speeding up
power spectrum and likelihood calculation, it is useful to design
efficient Markov Chain Monte Carlo samplers to do reliable
estimations with shorter chains.

We introduce the Hamiltonian Monte Carlo method (HMC) to improve the
efficiency of the MCMC methods used in Bayesian parameter estimation
in Astrophysics. This technique is based on Hamiltonian dynamics
instead of the random walk. Hamiltonian dynamics allows the chain to
move along trajectories of constant energy, taking large jumps in
the parameter space with relatively inexpensive computations. This
improves the acceptance rate and at the same time decreases the
correlations among the chains. Unlike the traditional MCMC methods,
the efficiency of the HMC remains constant even at large dimensions.
Therefor shorter chains will be needed for a reliable parameter
estimation comparing to a traditional MCMC chain yielding the same
performance. Besides that, the HMC is well suited to sample curved,
multi-modal and non-Gaussian distributions which are very hard to
sample from using the traditional MCMC methods.

In addition to idealized toy models, we have applied this technique
to cosmological parameter estimation. Our results have been compared
to the outputs of the standard package of \texttt{CosmoMC}.  For the
6-parameter flat LCDM model, the HMC is almost a factor of $10$ more
efficient than the default MCMC method of \texttt{CosmoMC}.
Therefore, using the HMC, one needs a much shorter chain to obtain
the same accuracy compared to the Metropolis method.

In addition to the high efficiency, another important feature of the
HMC is that we don't have to compromise on the accuracy of our
calculations. No approximation is made and the HMC sampler
faithfully samples from the target distribution.  It is shown in
this paper that even a rough estimate of the gradient of the
logarithm of likelihood can be used to make the HMC work. But the
better the gradients are estimated the higher the acceptance rate
will be. Combining this with methods of speeding up power spectrum
and likelihood calculation such as \texttt{Pico} will make the
Bayesian parameter estimation unbelievably fast.

\begin{acknowledgments}
I would like to thank David Spergel for encouraging me to work on this problem and for comments on the draft. Modifying Lico was made possible with the kind help of Chad Fendt.I wish to thank him and Joanna Dunkley, Arthur Kosowsky, Licia Verde and Carl E. Rasmussen for enlightening discussions. I also thank Neil Cornish for his helpful comments on the early days of this work and Antony Lewis and Benjamin Wandelt for their valuable comments on the draft.  Some of the
results in this paper have used the diagnostic tools of the
\texttt{MCMCstuff} package \cite{MCMCstuff}. The author acknowledges support
from NASA grant LTSA03-0000-0090.
\end{acknowledgments}

\end{document}